\documentclass[aps, prd, floatfix, nofootinbib, showpacs, twocolumn, superscriptaddress]{revtex4-1}

\usepackage{amsmath,amsfonts,amssymb,bm}
\usepackage{graphicx}
\usepackage{color}
\usepackage{enumitem}
\usepackage{multirow}
\usepackage{subfigure}
\usepackage{slashed}
\usepackage{textcomp}

\usepackage[mathscr,scaled=1.15]{urwchancal}
\DeclareFontFamily{OT1}{pzc}{}
\DeclareFontShape{OT1}{pzc}{m}{it}%
{<-> s * [1.15] pzcmi7t}{}
\DeclareMathAlphabet{\mathpzc}{OT1}{pzc}{m}{it}

\definecolor{purple}{rgb}{0.5,0,0.5}
\definecolor{blue}{rgb}{0.0,0,0.9}
\definecolor{prdblue}{rgb}{0.133,0.118,0.498}
\usepackage[colorlinks=true, pdfstartview=FitV, linkcolor=prdblue, citecolor= prdblue, urlcolor=prdblue]{hyperref}

\begin{document}


\title{$\,$\\[-7ex]\hspace*{\fill}{\normalsize{\sf\emph{Preprint no}. NJU-INP 014/19}}\\[1ex]
Effective charge from lattice QCD}

\author{Z.-F.~Cui}
\affiliation{School of Physics, Nanjing University, Nanjing, Jiangsu 210093, China}
\affiliation{Institute for Nonperturbative Physics, Nanjing University, Nanjing, Jiangsu 210093, China}
\author{J.-L.~Zhang}
\affiliation{Department of Physics, Nanjing Normal University, Nanjing 210023, China}
\author{D.~Binosi}
\affiliation{European Centre for Theoretical Studies in Nuclear Physics
and Related Areas; Villa Tambosi, Strada delle Tabarelle 286, I-38123 Villazzano (TN), Italy}
\author{F.~De Soto}
\affiliation{Dpto. Sistemas F\'isicos, Qu\'imicos y Naturales, Univ.\ Pablo de Olavide, E-41013 Sevilla, Spain}
\author{C.~Mezrag}
\affiliation{IRFU, CEA, Universit\'e Paris-Saclay, F-91191 Gif-sur-Yvette, France}
\author{J.~Papavassiliou}
\affiliation{Department of Theoretical Physics and IFIC,
University of Valencia and CSIC, E-46100, Valencia, Spain}
\author{C.~D.~Roberts}
\email[]{cdroberts@nju.edu.cn}
\affiliation{School of Physics, Nanjing University, Nanjing, Jiangsu 210093, China}
\affiliation{Institute for Nonperturbative Physics, Nanjing University, Nanjing, Jiangsu 210093, China}
\author{J.~Rodr\'{\i}guez-Quintero}
\email[]{jose.rodriguez@dfaie.uhu.es}
\affiliation{Department of Integrated Sciences and Center for Advanced Studies in Physics, Mathematics and Computation;
University of Huelva, E-21071 Huelva; Spain.}
\author{J.~Segovia}
\affiliation{Dpto. Sistemas F\'isicos, Qu\'imicos y Naturales, Univ.\ Pablo de Olavide, E-41013 Sevilla, Spain}
\affiliation{Institute for Nonperturbative Physics, Nanjing University, Nanjing, Jiangsu 210093, China}
\author{S.~Zafeiropoulos}
\affiliation{Aix Marseille Univ, Université de Toulon, CNRS, CPT, F-13288 Marseille, France}

\date{17 December 2019}

\begin{abstract}
Using lattice configurations for quantum chromodynamics (QCD) generated with three domain-wall fermions at a physical pion mass, we obtain a parameter-free prediction of QCD's renormalisation-group-invariant process-independent effective charge, $\hat\alpha(k^2)$.  Owing to the dynamical breaking of scale invariance, evident in the emergence of a gluon mass-scale, $m_0= 0.43(1)\,$GeV, this coupling saturates at infrared momenta: $\hat\alpha(0)/\pi=0.97(4)$.  Amongst other things: $\hat\alpha(k^2)$ is almost identical to the process-dependent (PD) effective charge defined via the Bjorken sum rule; and also that PD charge which, employed in the one-loop evolution equations, delivers agreement between pion parton distribution functions computed at the hadronic scale and experiment.  The diversity of unifying roles played by $\hat\alpha(k^2)$ suggests that it is a strong candidate for that object which represents the interaction strength in QCD at any given momentum scale; and its properties support a conclusion that QCD is a mathematically well-defined quantum field theory in four dimensions.
\end{abstract}




\maketitle


\section{Introduction}
QCD fascinates for many reasons, with the feature of confinement looming large amongst them.  At issue here is the definition.  When communicating about confinement, a typical practitioner has a notion in mind; yet the perspectives of any two different practitioners are often distinct, \emph{e.g}.\ Refs.\,\cite{Wilson:1974sk, Gribov:1998kb, Cornwall:1981zr}.  The proof of one expression of confinement will be contained within a demonstration that quantum $SU_c(3)$ gauge field theory is mathematically well-defined, \emph{i.e}.\ a solution to the ``Millennium Problem'' \cite{millennium:2006}.  However, that may be of limited value because Nature has provided light-quark degrees-of-freedom, which seemingly play a crucial r\^ole in the empirical realisation of confinement, perhaps because they enable screening of colour charge at low coupling strengths \cite{Gribov:1998kb}.

The QCD running coupling lies at the heart of many attempts to define and understand confinement because, almost immediately following the demonstration of asymptotic freedom \cite{Politzer:2005kc, Wilczek:2005az, Gross:2005kv}, the associated appearance of an infrared Landau pole in the perturbative expression for the running coupling spawned the idea of infrared slavery, \emph{viz}.\ confinement expressed through a far-infrared divergence in the running coupling.  In the absence of a nonperturbative definition of a unique running coupling, this idea is not more than a conjecture.  Notwithstanding that, and possibly inspired by the challenge, attempts to solve the confinement puzzle by completing the nonperturbative definition and calculation of a running coupling in QCD have received ongoing attention. (See Refs.\,\cite{Dokshitzer:1998nz, Grunberg:1982fw} and citations thereof.)

The archetypal running coupling is that computed in quantum electrodynamics (QED) more than sixty years ago \cite{GellMann:1954fq}: it is now known to great accuracy \cite{Tanabashi:2018oca} and the running has been observed directly \cite{Odaka:1998ui, Mele:2006ji}.  This Gell-Mann--Low effective charge is a renormalisation group invariant (RGI) and process-independent (PI) running coupling, which is obtained simply by computing the photon vacuum polarisation.  That is because ghost fields decouple in Abelian theories;  hence, one has the Ward identity \cite{Ward:1950xp}, which guarantees that the electric-charge renormalisation constant is equivalent to that of the photon field.  Stated physically, the impact of dressing the interaction vertices is absorbed into the vacuum polarisation.

This is not usually true in QCD because ghost fields do not decouple; consequently, the renormalisation constants associated with the running coupling and the gluon vacuum polarisation are different.  However, there is one approach to analysing QCD's Schwinger functions that preserves some of QED's simplicity; namely, the combination of pinch technique (PT) \cite{Cornwall:1981zr, Cornwall:1989gv, Pilaftsis:1996fh, Binosi:2009qm} and background field method (BFM) \cite{Abbott:1980hw}.   This framework acts to make QCD ``look'' Abelian: one systematically rearranges classes of diagrams and their sums in order to obtain modified Schwinger functions that satisfy linear Slavnov-Taylor identities \cite{Taylor:1971ff, Slavnov:1972fg}.  In the gauge sector, using Landau gauge, this produces a modified gluon dressing function from which one can compute a unique QCD running coupling, \emph{i.e}.\ the PT-BFM polarisation captures all required features of the renormalisation group.  Furthermore, the coupling is process independent: one obtains precisely the same result, independent of the scattering process considered, whether gluon+gluon$\,\to\,$gluon+gluon, quark+quark$\,\to\,$quark+quark, \emph{etc}.  This clean connection between the coupling and the gluon vacuum polarisation relies on another particular feature of QCD, \emph{viz}.\ in Landau gauge the renormalisation constant of the gluon-ghost vertex is unity \cite{Taylor:1971ff}, in consequence of which the effective charge obtained from the PT-BFM gluon vacuum polarisation is directly connected with that deduced from the gluon-ghost vertex \cite{Sternbeck:2007br, Boucaud:2008gn, Aguilar:2009nf}, sometimes called the ``Taylor coupling,'' $\alpha_{\rm T}$ \cite{Blossier:2011tf, Blossier:2012ef}.

These observations underly the RGI PI effective coupling, $\hat\alpha(k^2)$, introduced in Ref.\,\cite{Binosi:2016nme}.  Therein, a combination of continuum- and lattice-QCD methods was used to complete the first calculation of $\hat\alpha(k^2)$, which was subsequently refined \cite{Rodriguez-Quintero:2018wma}.  With improvements in lattice configurations, it is worth returning to that effective charge.  In Sec.\,\ref{SecPIEC} we recapitulate the discussion in Ref.\,\cite{Binosi:2016nme}.  Then, in Sec.\,\ref{SecNewPrediction}, using the most up-to-date lattice-QCD (lQCD) configurations available, we update the prediction for $\hat\alpha(k^2)$ and comment upon its relevance to confinement and the nonperturbative definition of QCD.  Section\,\ref{SecPDEC} draws novel connections between $\hat\alpha(k^2)$ and an often discussed process-dependent effective charge \cite{Deur:2005cf, Deur:2008rf, Brodsky:2010ur}, highlighting the possibility that $\hat\alpha(k^2)$ may provide an objective measure of the interaction strength in QCD at any given momentum scale.  Section~\ref{epilogue} presents a summary and offers perspectives.


\section{PI effective coupling}
\label{SecPIEC}
We begin with the mathematical foundation provided by Refs.\,\cite{Binosi:2014aea, Binosi:2016xxu}; namely, using the PT-BFM method, one derives the following identities $\!(T_{\mu\nu}(k)=\delta_{\mu\nu}-k_\mu k_\nu/k^2)$:
\begin{subequations}
\label{allhatd}
\begin{align}
\alpha(\zeta^2) & D^{\rm PB}_{\mu\nu}(k;\zeta)
%
%
\label{hatd}
 = \widehat{d}(k^2)\, T_{\mu\nu}(k)\,, \\
%
%
  {\mathpzc I}(k^2) &:= k^2 \widehat{d}(k^2)=
  \frac{\alpha_{\rm T}(k^2) }
  { [1-L(k^2;\zeta^2)F(k^2;\zeta^2)]^2}\,,
  \label{mathcalI}\\
\alpha_{\rm T}(k^2) & = \alpha(\zeta^2) k^2 \Delta(k^2;\zeta^2) F^2(k^2;\zeta^2) \,,
\end{align}
\end{subequations}
where:
\begin{enumerate}[label=(\roman*)]
\item $\widehat{d}(k^2)$ is the RGI PI running-interaction %
 discussed in Ref.\,\cite{Aguilar:2009nf};

\item $\alpha(\zeta^2):=g^2(\zeta^2)/[4\pi]$, where $g$ is the Lagrangian coupling and $\zeta$ the renormalisation scale;

\item $D^{\rm PB}_{\mu\nu}(k)= \Delta^{\rm PB}(k^2) T_{\mu\nu}(k)$ is the PT-BFM gluon two-point function;
\item $D_{\mu\nu}(k) = \Delta(k^2) T_{\mu\nu}(k)$ is the canonical gluon two-point function;

\item $F$ is the dressing function for the ghost propagator;
\item
and $L$ is that longitudinal part of the gluon-ghost vacuum polarisation which vanishes at $k^2=0$ \cite{Aguilar:2009nf}.
\end{enumerate}
(The RGI character of $\hat d(k^2)$ has explicitly been verified: numerically via direct calculation \cite{Binosi:2014aea} and analytically in the infrared and ultraviolet limits \cite{Binosi:2016xxu}.)

Using Eqs.\,\eqref{allhatd}, QCD's matter-sector gap equation -- the dressed-quark Dyson-Schwinger equation (DSE) -- can be written $(k=p-q)$
\begin{subequations}
\label{gendseN}
\begin{align}
S^{-1}(p) 
& = Z_2 \,(i\gamma\cdot p + m^{\rm bm}) + \Sigma(p)\,,\\
\Sigma(p)& =  Z_2\int^\Lambda_{dq}\!\!
4\pi \widehat{d}(k^2) \,T_{\mu\nu}(k)\gamma_\mu S(q) \hat\Gamma^a_\nu(q,p)\, ,
\end{align}
\end{subequations}
where $\int_{dq}^\Lambda$ represents a Poincar{\'e} invariant regularisation of the four-dimensional integral, with $\Lambda$ the regularisation mass-scale;
and the usual $Z_1 \Gamma^a_\nu$ has become $Z_2 \hat\Gamma^a_\nu$, with the latter function being a PT-BFM gluon-quark vertex that satisfies an Abelian-like Ward-Green-Takahashi identity \cite{Binosi:2009qm} and $Z_{1,2}$ are, respectively, the gluon-quark vertex and quark wave function renormalisation constants.

Here it is useful to review the ultraviolet and infrared limits of the RGI interaction, $\widehat{d}(k^2)$, and draw its connection with the PI effective charge.

At momenta far above $\Lambda_{\rm QCD}$, the mass-scale characterising QCD perturbation theory,\footnote{Renormalising via momentum-subtraction, $\Lambda_{\rm QCD} = 0.58\,$GeV when three quark flavours are active \protect{\cite{Zafeiropoulos:2019flq}}.}
one has \cite{Binosi:2016xxu}
\begin{equation}
L(k^2;\zeta^2)F(k^2;\zeta^2) \underset{k^2\gg \Lambda_{\rm QCD}^2}{\approx} \frac{3}{8\pi} \alpha(k^2) +
{\cal O}\left( \alpha^2 \right)  \, .
\label{eq:LFinIR}
\end{equation}
It is not necessary to specify a renormalisation scheme because all couplings are equivalent at one-loop level.
Hence, ${\mathpzc I}(k^2)$ in Eq.\,\eqref{mathcalI} is a PI running coupling at ultraviolet momenta:
\begin{subequations}
\label{eq:PIpert}
\begin{align}
{\mathpzc I}(k^2) &= \frac{\alpha_{\rm T}(k^2)}{[1-L(k^2;\zeta^2)F(k^2;\zeta^2)]^2}
 \\
&=  \alpha_T(k^2) \left[ 1 + \frac 3 {4 \pi} \alpha_T(k^2) + {\cal O}\left(\alpha^2_T\right) \right]\, .
\end{align}
\end{subequations}

At the other extreme: $k^2\ll \Lambda_{\rm QCD}^2$, one encounters a signature feature of strong QCD; namely, $\widehat{d}(k^2)$ saturates to a finite value at $k^2=0$ owing to the nonperturbative generation of a mass-scale in the gauge sector, \emph{e.g}.\ Refs.\,\cite{Cucchieri:2007rg, Aguilar:2008xm, Dudal:2008sp, Bogolubsky:2009dc, Boucaud:2008ky, Aguilar:2012rz, Strauss:2012dg, Cyrol:2017ewj, Pelaez:2017bhh, Siringo:2018uho, Binosi:2019ecz}.  In fact \cite{Aguilar:2009nf, Binosi:2014aea}
\begin{subequations}
\label{dhatagain}
\begin{align}
0< \hat d(k^2=0) & =\alpha(\zeta^2)\Delta^{\rm PB}(k^2=0;\zeta^2)  \\
& = \frac{\alpha(\zeta^2)}{\hat m_g^2(\zeta^2)} = \frac{\alpha_0}{m_0^2} \,.
\end{align}
\end{subequations}

It is useful to connect this outcome with the canonical gluon two-point function, which satisfies, using Eqs.\,\eqref{allhatd}:
\begin{align}
\Delta(k^2;\zeta^2) & F^2(k^2;\zeta^2)  = \nonumber \\
& \Delta^{\rm PB}(k^2;\zeta^2) [1-L(k^2;\zeta^2)F(k^2;\zeta^2)]^2.
\end{align}
One can write
\begin{align}
%
\label{DeltamCan}
\Delta^{-1}(k^2;\zeta^2) &= k^2 J(k^2;\zeta^2) + m^2_g(k^2;\zeta^2) \, ,
\end{align}
where $m_g(k^2,\zeta)$ is the canonical dynamical gluon mass function and $J(k^2,\zeta)$ is the associated kinetic term.  For later use, we note \cite{Aguilar:2013vaa, Athenodorou:2016oyh}:
\begin{subequations}
\label{Jprops}
\begin{align}
J(k^2,\zeta) & \stackrel{k^2\ll \Lambda_{\rm QCD}^2}{\sim} \ln(k^2) + {\cal O}(k^2)\,, \\
J(k^2,\zeta) & \stackrel{k^2\gg \Lambda_{\rm QCD}^2}{\sim}
[\ln(k^2/\Lambda_{\rm QCD}^2)/\ln(\zeta^2/\Lambda_{\rm QCD}^2)]^{\gamma_0/\beta_0} \nonumber \\
& \qquad \qquad \times  (1 + {\cal O}(1/k^2))\,,
\end{align}
\end{subequations}
with $\gamma_0= 13/2 - 2 n_f /3$, $\beta_0 = 11 - 2 n_f /3$, $n_f$ is the number of active quark flavours.  Now \begin{align}
\hat m_g(\zeta^2) & = m_g(0;\zeta^2) / F(0;\zeta^2)
\end{align}
follows from Eqs.\,\eqref{dhatagain}\,--\,\eqref{DeltamCan}.

Consider the product
\begin{equation}
\label{CalDDelta}
{\mathsf D}(k^2) = \Delta(k^2;\zeta^2) \, m_g^2(0;\zeta^2) / m_0^2,
\end{equation}
which is a RGI function.  Now use Eqs.\,\eqref{Jprops} to develop an interpolation, ${\mathpzc D}(k^2)$, which accurately describes available results for $\mathsf D(k^2)$ on $k^2 \lesssim \zeta^2$ and yet also expresses the following behaviour:
\begin{equation}
\label{CalDlimits}
\frac{1}{{\mathpzc D}(k^2)} =
\left\{\begin{array}{ll}
m_0^2 + {\rm O}(k^2 \ln k^2) & k^2 \ll \Lambda_{\rm QCD}^2\\
\rule{0ex}{3ex}
k^2 + {\rm O}(1) & k^2 \gg \Lambda_{\rm QCD}^2\\
\end{array}\right. ,
\end{equation}
\emph{i.e}.\ in both the far-infrared and -ultraviolet, ${\mathpzc D}(k^2)$ behaves as the free propagator for a boson with mass $m_0$.  Then, writing
\begin{equation}
\label{Definedhat}
\hat d(k^2) = \hat\alpha(k^2)\, {\mathpzc D}(k^2)\,,
\end{equation}
one arrives at an effective charge, $\hat\alpha(k^2)$, which is:
\begin{enumerate}[label=(\alph*)]
\item RGI;
\item PI, hence, key to the unified description of an extensive array of hadron observables;
\item identical to the standard QCD running-coupling in the ultraviolet, Eq.\,\eqref{eq:PIpert};
\item saturates to a finite value, $\alpha_0$, on $k^2\ll \Lambda_{\rm QCD}^2$;
\end{enumerate}
and completely determined by those functions from QCD's gauge sector which connect the canonical and PT-BFM gluon two-point functions.  Concretely expressed,
\begin{subequations}
\label{alphahatequation}
\begin{align}
\hat\alpha(k^2) & = \alpha_0 \frac{{\mathsf D}(k^2)}{{\mathpzc D}(k^2)}\left[\frac{F(k^2;\zeta^2)/F(0;\zeta^2)}{1-L(k^2;\zeta^2)F(k^2;\zeta^2)}\right]^2 \\
& \stackrel{k^2\lesssim \zeta^2}{=} \alpha_0 \left[\frac{F(k^2;\zeta^2)/F(0;\zeta^2)}{1-L(k^2;\zeta^2)F(k^2;\zeta^2)}\right]^2.
\end{align}
\end{subequations}
Crucially, each of these functions can be computed using continuum and/or lattice methods.

In terms of this PI charge, the quark DSE, Eq.\,\eqref{gendseN}, can be rewritten:
\begin{subequations}
\label{gendseNew}
\begin{align}
S^{-1}(p) 
& = Z_2 \,(i\gamma\cdot p + m^{\rm bm}) + \Sigma(p)\,,\\
\Sigma(p)& =  Z_2\int^\Lambda_{dq}\!\!
4\pi \hat\alpha(k^2) {\mathpzc D}_{\mu\nu}(k) \gamma_\mu S(q) \hat\Gamma^a_\nu(q,p)\, ,
\end{align}
\end{subequations}
where ${\mathpzc D}_{\mu\nu}(k) = {\mathpzc D}(k^2) T_{\mu\nu}(k)$.  This series of observations enabled unification of that body of work directed at the \emph{ab initio} computation of QCD's effective interaction via direct analyses of gauge-sector gap equations and the studies which aimed to infer the interaction by fitting data within a symmetry-preserving truncation of those equations in the matter sector that are relevant to bound-state properties \cite{Binosi:2014aea}.

It is worth remarking here that $\hat{\alpha}(k^2)$ is RGI and PI in any gauge.  Moreover, it is sufficient to calculate this charge in Landau gauge because:
$\hat{\alpha}(k^2)$ is form-invariant under gauge transformations, since the identities expressed by Eqs.\,\eqref{allhatd} are the same in all linear covariant gauges \cite{Binosi:2013cea};
and gauge covariance ensures that such transformations are implemented by multiplying a simple factor into the configuration space transform of the gap equation's solution and may consequently be absorbed into the dressed-quark two-point function \cite{Aslam:2015nia}.

\section{Effective Charge from Modern Lattice Simulations}
\label{SecNewPrediction}
\subsection{Existing Results}
Existing analyses of continuum and lattice results for QCD's gauge sector yield the PI coupling depicted as the solid black curve in Fig.\,2 of Ref.\,\cite{Rodriguez-Quintero:2018wma}, corresponding to $\alpha_0/\pi = 1.00$, $m_0 = 0.47\,$GeV$\approx m_p/2$, where $m_p$ is the proton mass.  These results were based on lQCD simulations obtained with four dynamical flavours of twisted-mass fermions \cite{Blossier:2011tf, Blossier:2012ef} at pion masses $m_\pi \gtrsim 0.3\,$GeV.

Today, new simulations exist with three domain-wall fermions and $m_\pi = 0.139\,$GeV.  They were recently employed \cite{Zafeiropoulos:2019flq} to compute the Taylor coupling at intermediate and large momenta; and therefrom deduce the $\overline{\rm MS}$-coupling at the $Z^0$ mass, producing a result in agreement with the world average \cite{Tanabashi:2018oca}.  We now take advantage of the gauge-sector two-point Schwinger functions computed with these state-of-the-art lattice configurations in order to deliver a refined prediction for $\hat\alpha(k^2)$.

\subsection{Gluon two-point function}
\begin{figure}[t]
\begin{tabular}{c}
	\includegraphics[width=0.9\linewidth]{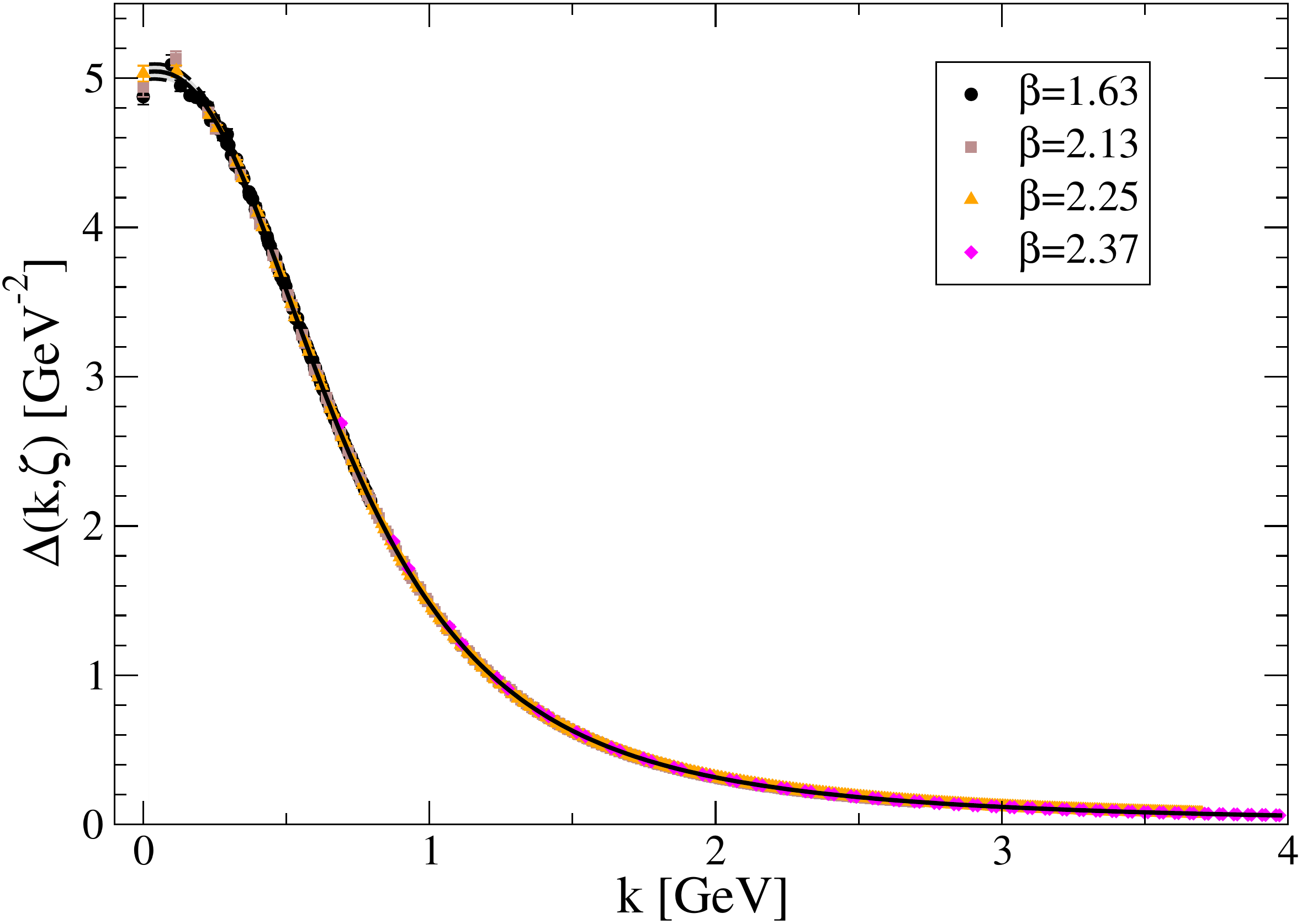} \\
	\includegraphics[width=0.92\linewidth]{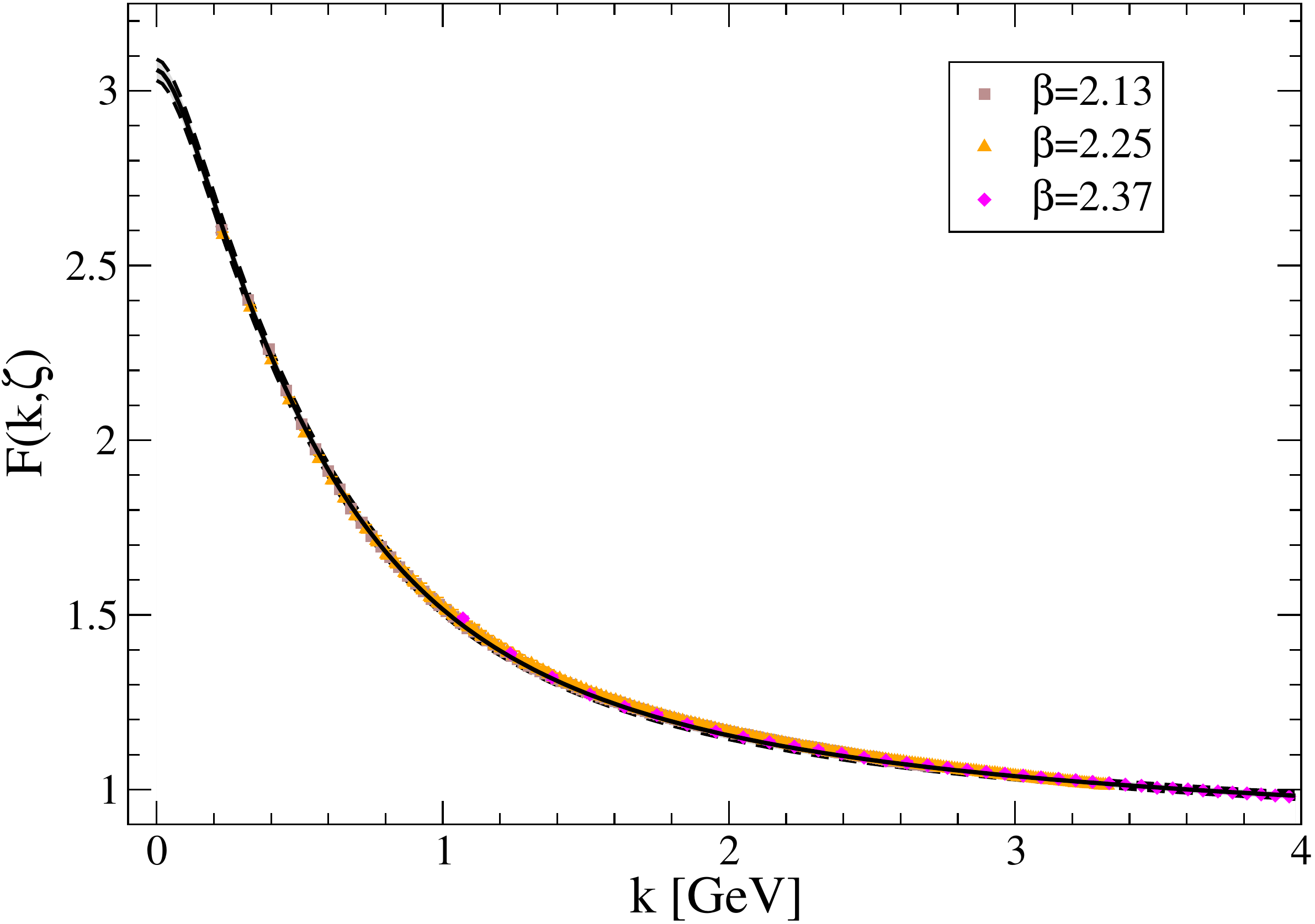}
\end{tabular}
\caption{\label{fig:props}
Two-point Schwinger functions -- gluon (upper panel) and ghost  (lower) -- obtained from lQCD simulations: gauge field ensembles with $n_f=3$ domain-wall fermions \cite{Blum:2014tka, Boyle:2015exm, Boyle:2017jwu} at the physical pion mass.  (Renormalisation point: $\zeta = 3.6\,$GeV.)
Details of the lattice configurations can be found in Ref.\,\cite{Zafeiropoulos:2019flq}-Table~I,
save for those at $\beta=1.63$.  In this case, all features are as described in connection with Ref.\,\cite{Boyle:2015exm}-Table~II, except the volume, which is $48^3\times 64$.
Solid black curve: \emph{upper panel} -- least-squares fit to lattice output defined by Eqs.\,\eqref{DeltaFit}, \eqref{DeltaParams};
and \emph{lower panel} -- solution of the ghost DSE described in Sec.\,\ref{SecGhostDSE}.  (Uncertainty bands discussed in Sec.\,\ref{SecEffectiveCharge}.)
}
\end{figure}

The gluon two-point function is obtained via Monte-Carlo averaging over gauge-field lattice configurations constrained to Landau gauge \cite{Ayala:2012pb, Binosi:2016xxu}.  Using the MOM renormalisation scheme with $\zeta = 3.6\,$GeV, the configurations from Ref.\,\cite{Blum:2014tka, Boyle:2015exm, Boyle:2017jwu} yield the results for $\Delta(k)$ depicted in the upper panel of Fig.\,\ref{fig:props}.  ($\zeta = 3.6\,$GeV is chosen because it lies within both the domain of reliable lattice output and perturbative-QCD validity.)

The information required herein is best obtained by developing an accurate interpolation of the lattice results in Fig.\,\ref{fig:props}-upper panel.  Informed by Eqs.\,\eqref{DeltamCan}, \eqref{Jprops}, as discussed in connection with Eqs.\,\eqref{CalDDelta}, \eqref{CalDlimits}, we use ($s=k^2$)
\begin{equation}
\label{DeltaFit}
\Delta^{\rm fit}(s;\zeta^2) =
\frac{\lambda_0(\zeta^2)[1 + n_1 s]}
{1 - \delta s \ln[1+m_0^2/s] + d_1 s + d_2^2 s^2 \ell(s)}\,,
\end{equation}
where
$\ell(s) = [\ln([s+\zeta^2]/\Lambda_{\rm QCD}^2)/\ln(\zeta^2/\Lambda_{\rm QCD}^2)]^{\gamma_0/\beta_0}$
and all dependence on the renormalisation scale is otherwise carried by $\lambda_0(\zeta^2)$.
Using this form, a least-squares fit yields (all quantities in units ${\rm GeV}^{-2}$):
\begin{equation}
\label{DeltaParams}
\begin{array}{ccccc}
\lambda_0 & n_1 & \delta & d_1 & d_2 \\
5.041 & 0.282 & 0.432 & 1.887 & 1.241
\end{array}\,.
\end{equation}
The fitting function is drawn as the solid curve in Fig.\,\ref{fig:props}-upper panel: it provides an excellent description of the lattice output.  Cross-referencing with Eq.\,\eqref{DeltamCan},  
and, subsequently, Eqs.\,\eqref{CalDDelta}, \eqref{CalDlimits}, one obtains (in GeV)
\begin{equation}
\label{DeltaAnswers}
m_g(0;\zeta^2) = \frac{1}{\lambda_0^{1/2}} = 0.445 \,,
\quad
m_0 = \frac{n_1^{1/2}}{d_2} = 0.428\,.
\end{equation}
(Estimates of sensitivity to errors on inputs are provided in Sec.\,\ref{SecEffectiveCharge}.)


\subsection{Ghost dressing function and \mbox{$\mathbf{L(k^2;\zeta^2)}$}}
\label{SecGhostDSE}
The ghost propagator dressing function, $F(k^2;\zeta^2)$, can be obtained via inversion of the Faddeev-Popov operator \cite{Boucaud:2005gg, Ayala:2012pb, Binosi:2016xxu}.  Once again using the configurations in Refs.\,\cite{Blum:2014tka, Boyle:2015exm, Boyle:2017jwu} and MOM renormalisation at $\zeta=3.6\,$GeV, we obtain the points in Fig.\,\ref{fig:props}-lower panel.\footnote{Our inversion algorithm was ill-adapted to the $\beta=1.63$ ensemble, which has the largest volume but an awkward geometry: $L^3\times T$, $L=48$, $T=64$, $L/T = 3/4$, whereas $L/T=1/2$ for all other configurations.  Hence, $\beta=1.63$ results are not reported.}

The ghost dressing function can also be obtained by solving the associated gap equation $(p=k+q)$:
\begin{align}
 & F^{-1}(k^2;\zeta^2)  = \tilde Z_3(\zeta^2,\Lambda^2)  - 3 g^2(\zeta^2) \nonumber \\
&
\times \int_{dq}^\Lambda \left[1 - \frac{k \cdot q}{k^2 q^2}\right]
H_1(q,p) \Delta(q^2;\zeta^2) \frac{F(p^2;\zeta^2)}{p^2} \, ,
\label{eq:ghostgap}
\end{align}
where $\tilde Z_3$ is the ghost wave function renormalisation constant.
The function $H_1(q,p)$ in Eq.\,\eqref{eq:ghostgap} derives from the dressed ghost-gluon vertex.  In Landau gauge, $H_1(q,p)$ only differs modestly from unity \cite{Ayala:2012pb, Dudal:2012zx, Aguilar:2013xqa}.  Herein, we use the parametrisation of Ref.\,\cite[Eq.\,(4.6)]{Ayala:2012pb}, which incorporates all infrared physics that contributes materially to the solution of Eq.\,\eqref{eq:ghostgap}.  Stated differently, the parametrisation enables one to obtain a solution of Eq.\,\eqref{eq:ghostgap} in excellent agreement with available lattice points, as evidenced by the solid black curve in Fig.\,\ref{fig:props}-lower panel.
More complicated parametrisations are available \cite{Boucaud:2011eh}, but they do not improve the computed result for $F(k^2;\zeta^2)$.
In solving Eq.\,\eqref{eq:ghostgap}, we use Eqs.\,\eqref{DeltaFit}, \eqref{DeltaParams} for the gluon two-point function, $g^2(\zeta^2)=4.44$ at $\zeta=3.6\,$GeV \cite{Zafeiropoulos:2019flq}, and choose $\tilde Z_3(\zeta^2;\Lambda^2)$ such that $F(\zeta^2;\zeta^2)=1$.

\begin{figure}[t]
\begin{tabular}{c}
	\includegraphics[width=0.9\linewidth]{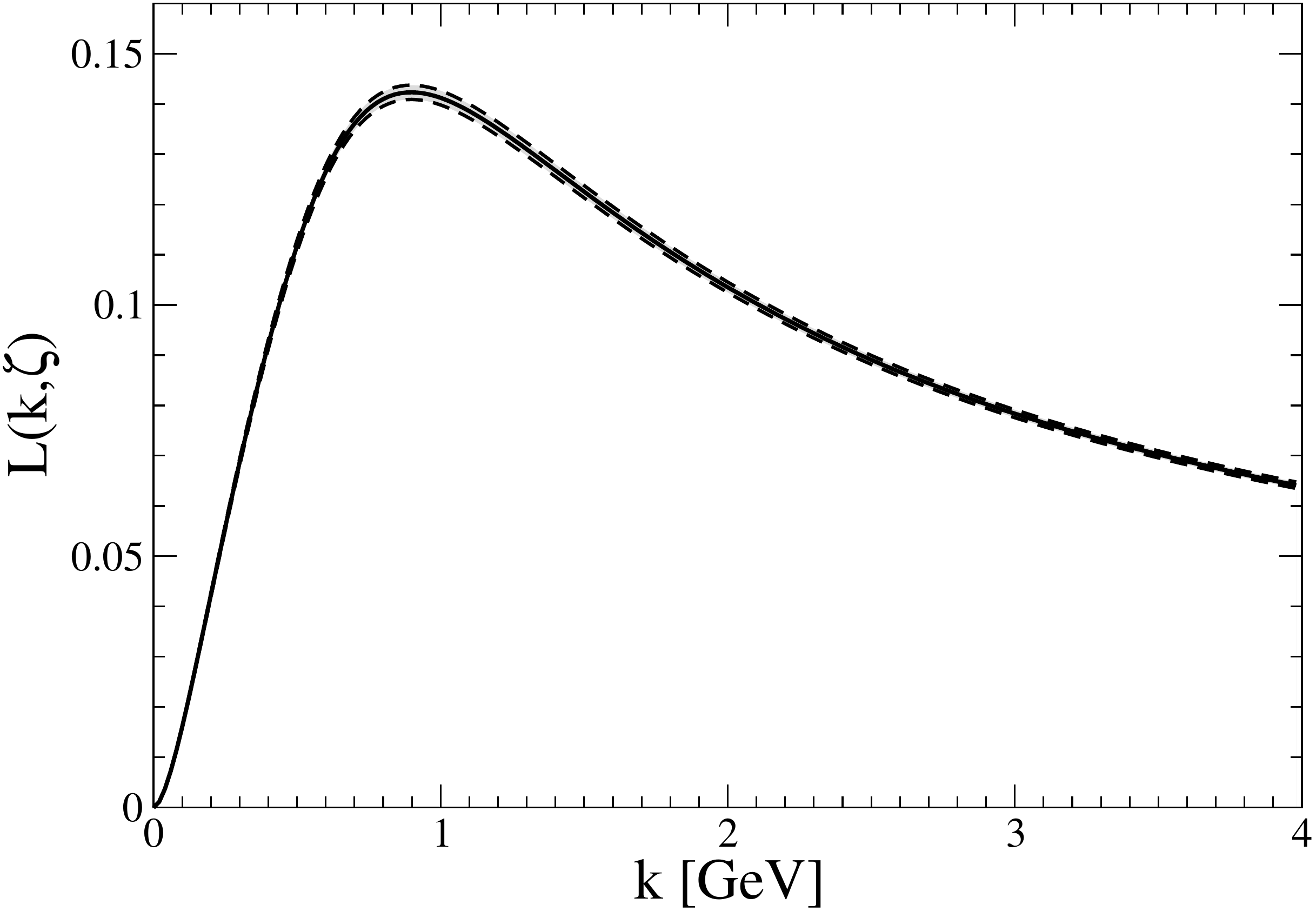} \\
	\includegraphics[width=0.9\linewidth]{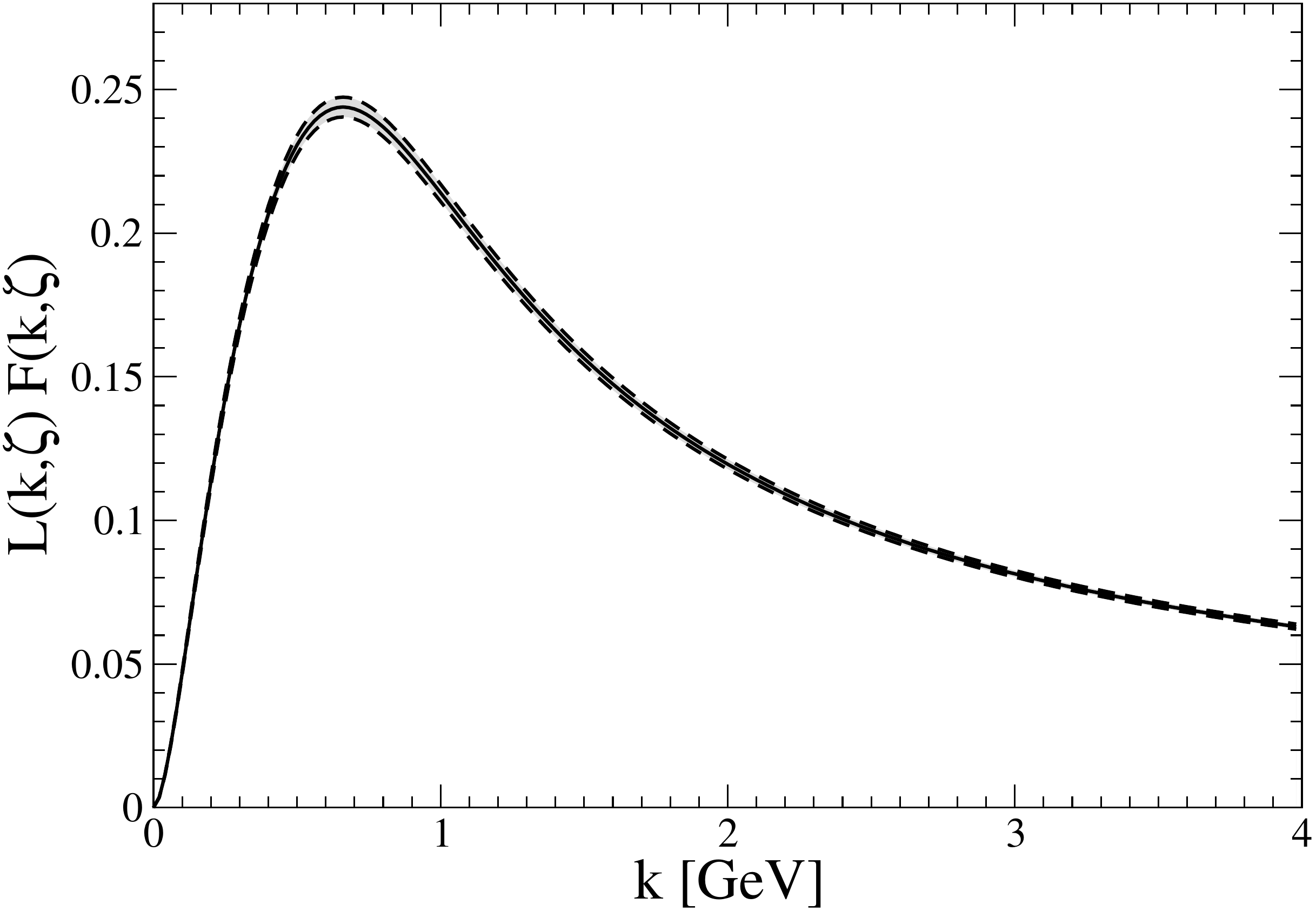}
\end{tabular}
\caption{\label{fig:LFk}
\emph{Upper panel} -- $L(k^2;\zeta^2)$ obtained from Eq.\,\eqref{eq:Lk}.
\emph{Lower panel} -- product $L(k^2;\zeta^2) F(k^2;\zeta^2)$ that appears in computing $\hat d(k^2)$; and hence, $\hat \alpha(k^2)$.}
\end{figure}

It remains only to compute $L(k^2;\zeta^2)$, the longitudinal part of the gluon-ghost vacuum polarisation.  This function vanishes at $k^2=0$ and is perturbatively small on $k^2\gg \Lambda_{\rm QCD}^2$; but its behaviour at intermediate momenta has a significant effect on $\hat d(k^2)$ and, hence, the PI effective charge \cite{Binosi:2016nme}.  Following Ref.\,\cite{Aguilar:2009nf}:
\begin{align}
&L(k^2;\zeta^2)  = g^2(\zeta) \nonumber\\
& \times \int_{dq}^\Lambda \left[4 \frac{k \cdot q}{k^2 q^2} - 1 \right] H_1(q,p)
 \Delta(q^2;\zeta^2) \frac{F(p^2;\zeta^2)}{p^2} \, .
\label{eq:Lk}
\end{align}
As described above, every element in the integrand is accurately known; and using these inputs, one obtains the results in Fig.\,\ref{fig:LFk}.

\begin{figure}[t]
	\includegraphics[width=0.95\linewidth]{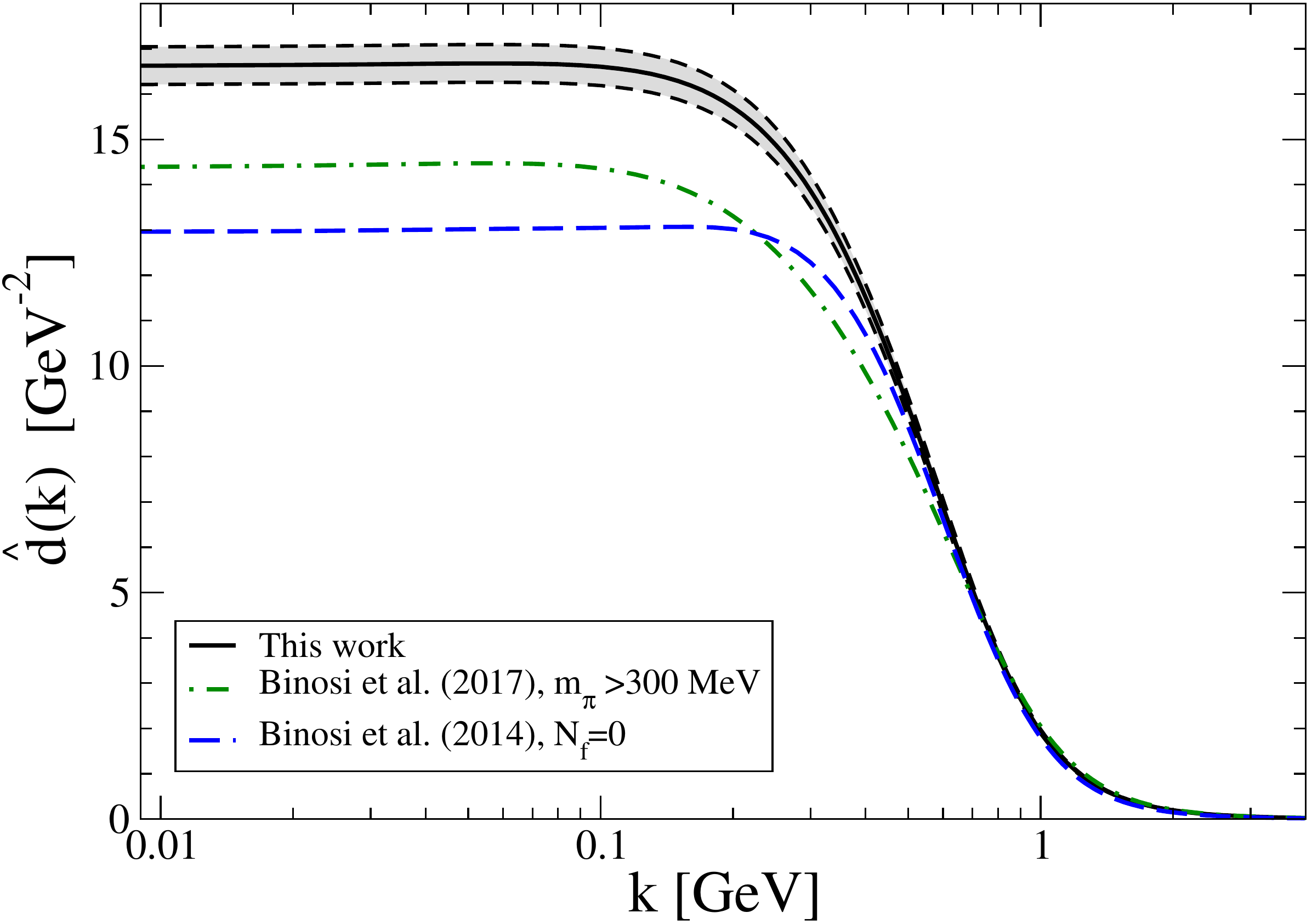} \\
\caption{\label{Figdhat}
$\hat d(k^2)$, RGI interaction reported in Ref.\,\cite{Binosi:2014aea} -- dashed blue curve;
in Ref.\,\cite{Binosi:2016xxu} -- dot-dashed green curve;
and computed herein -- solid black curve within bands.
The advances over time are largely driven by improvements in lQCD results for the gluon two-point function.
}
\end{figure}

\subsection{Effective Interaction}
\label{SecEffectiveCharge}
We now have all information required for calculation of $\hat d(k^2)$ and, hence, the RGI PI effective charge, $\hat \alpha(k^2)$.  However, before proceeding we provide an estimate of the sensitivity of our prediction to uncertainties on the inputs.
To that end, consider Fig.\,\ref{fig:props}-upper panel.  The results from all four lattices are in perfect agreement, except at the two lowest momenta.  At these positions, the deviation between our fit and the lattice points is $\lesssim 1$\%.  Naturally, low-momentum lattice results are suspect owing to finite-volume effects.  Acknowledging that, then by applying a uniform 1\% error to $\Delta^{\rm fit}(s)$ we express a conservative uncertainty estimate.
$\Delta^{\rm fit}(s)$ is the key element in the linear kernel of Eq.\,\eqref{eq:ghostgap}; hence, $F(k^2;\zeta^2)$ can also be uncertain at the level of 1\%.  The same is true for $L(k^2;\zeta^2)$.  (Given the precise agreement between our DSE solution and the lattice results for $F(k^2;\zeta^2)$, we neglect any error in $H_1$.)
Following this reasoning, we subsequently report results obtained by including a propagated 1\% uncertainty in each of these inputs.

In Fig.\,\ref{Figdhat} we depict our prediction for the RGI function $\hat d(k^2)$ defined in Eqs.\,\eqref{allhatd}.  It is characterised by the infrared value (in GeV$^{-2}$):
\begin{equation}
\hat d(k^2=0) = 16.6(4) \,;
\end{equation}
and in combination with Eqs.\,\eqref{dhatagain}, \eqref{DeltaFit}--\eqref{DeltaAnswers}, one obtains
\begin{equation}
\label{alpha0}
\alpha_0 = m_0^2 \, \hat d(k^2=0) = 0.97(4)\,\pi \,.
\end{equation}
Earlier analyses yielded \cite{Binosi:2014aea, Binosi:2016nme, Rodriguez-Quintero:2018wma}: $\alpha_0/\pi \approx 0.9\,$-$1.0$.  (N.B.\ Eq.\,\eqref{gendseNew} only supports emergence of a nonzero chiral-limit dressed-quark mass when $\alpha_0 \gtrsim 0.3\,\pi$ \cite{Roberts:1994dr}.)

In Fig.\,\ref{Figdhat}, we have also drawn the results for $\hat d(k^2)$ computed previously \cite{Binosi:2014aea, Binosi:2016xxu}: quantitative differences are evident.  The first calculation \cite{Binosi:2014aea} used quenched lQCD results for $\Delta(k^2;\zeta^2)$; the second \cite{Binosi:2016xxu} employed $n_f=2+1+1$ flavours of twisted-mass fermions with $m_\pi \gtrsim 0.3\,$GeV; and herein we used configurations built with $n_f=3$ flavours of domain-wall fermions and a physical pion mass.  It is this steady advance in lQCD results for the gluon two-point function that is behind our improved prediction for $\hat d(k^2)$.

\subsection{Effective Charge}
\label{SecEffectiveChargeReally}
\begin{figure}[t!]
\includegraphics[width=0.95\linewidth]{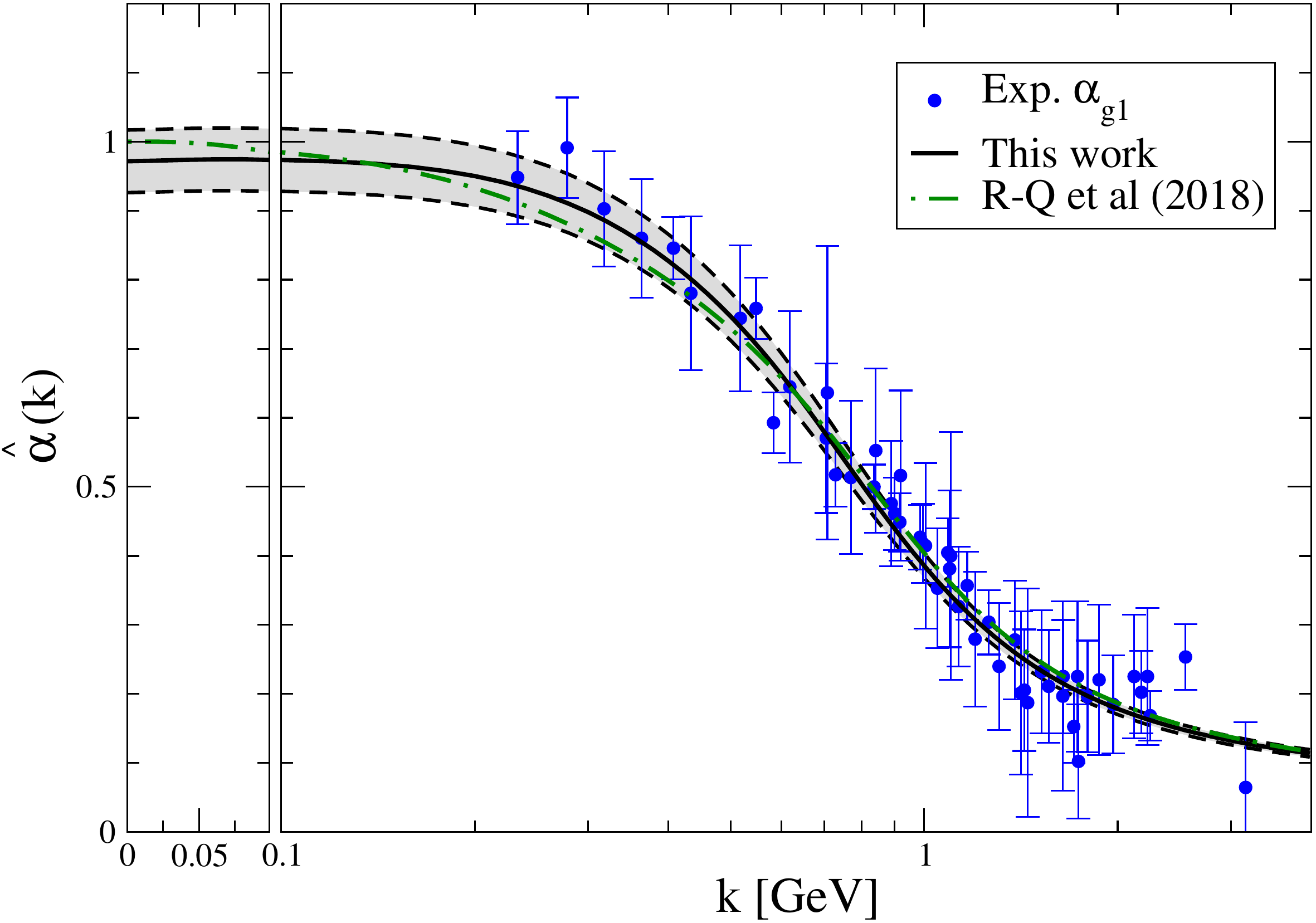}
\caption{\label{Figwidehatalpha}
Solid black curve within grey band -- RGI PI running-coupling $\hat{\alpha}(k^2)/\pi$, obtained herein using Eqs.\,\eqref{alphahatequation}, \eqref{alpha0}; and dot-dashed green curve -- earlier result (R-Q \emph{et al}.\ 2018) \cite{Rodriguez-Quintero:2018wma}.
For comparison, world data on the process-dependent charge, $\alpha_{g_1}$, is also depicted
\cite{%
Deur:2005cf, Deur:2008rf, Deur:2014vea,
Ackerstaff:1997ws, Ackerstaff:1998ja, Airapetian:1998wi, Airapetian:2002rw, Airapetian:2006vy,
Kim:1998kia,
Alexakhin:2006oza, Alekseev:2010hc, Adolph:2015saz,
Anthony:1993uf, Abe:1994cp, Abe:1995mt, Abe:1995dc, Abe:1995rn, Anthony:1996mw, Abe:1997cx, Abe:1997qk, Abe:1997dp, Abe:1998wq, Anthony:1999py, Anthony:1999rm, Anthony:2000fn, Anthony:2002hy}.
}
\end{figure}

The RGI PI effective charge is now available via Eqs.\,\eqref{alphahatequation}, \eqref{alpha0}; and the prediction deriving from the above analysis is drawn in Fig.\,\ref{Figwidehatalpha}.  Despite the differences evident in Fig.\,\ref{Figdhat}, the earlier predictions for $\hat\alpha(k^2)$ \cite{Binosi:2016nme, Rodriguez-Quintero:2018wma} lie within the grey shaded band, as illustrated using the result from Ref.\,\cite{Rodriguez-Quintero:2018wma}. This is because the differences apparent in Fig.\,\ref{Figdhat} owe largely to decreases in $m_0$ as the lQCD configurations have improved, modifications which leave $\alpha_0$ largely unchanged.

Figure\,\ref{Figwidehatalpha} shows that QCD's RGI PI coupling is everywhere finite, \emph{i.e}.\ there is no Landau pole and the theory likely possesses an infrared-stable fixed point.  The preceding discussion reveals that these features owe to the emergence of a nonzero mass-scale in QCD's gauge sector, something which must be an integral part of any solution to the $SU_c(3)$ ``Millennium Problem'' \cite{millennium:2006}.  Indeed, the fact that $m_0 \approx m_p/2$ indicates that the magnitude of scale-invariance violation in chiral-limit QCD is very large \cite{Roberts:2016vyn} and seems tuned to eliminate the Gribov ambiguity \cite{Gao:2017uox}.  Moreover, in concert with asymptotic freedom, such features support a view that QCD is unique amongst known four-dimensional quantum field theories in being defined and internally consistent at all energy scales.  This is intrinsically significant and might also have implications for attempts to develop an understanding of physics beyond the Standard Model based upon non-Abelian gauge theories \cite{Appelquist:1996dq, Sannino:2009za, Appelquist:2009ka, Hayakawa:2010yn, Cheng:2013eu, Aoki:2013xza, DeGrand:2015zxa, Binosi:2016xxu}.

\section{Process-Dependent Charge}
\label{SecPDEC}
Another approach to determining an ``effective charge'' in QCD was introduced in Ref.\,\cite{Grunberg:1982fw}.  This is a process-dependent procedure; namely, an effective running coupling is defined to be completely fixed by the leading-order term in the perturbative expansion of a given observable in terms of the canonical perturbative running coupling.  A potential issue with such a scheme is the process-dependence itself.  In principle, effective charges from different observables can be algebraically related via an expansion of one coupling in terms of the other.  However, expansions of this type would typically contain infinitely many terms \cite{Deur:2016tte}; consequently, the connection does not readily imbue a given process-dependent charge with the ability to predict any other observable, since the expansion is only defined \emph{a posteriori}, \emph{i.e}.\ after both effective charges are independently constructed.

One example is the process-dependent effective charge $\alpha_{g_1}(k^2)$, defined via the Bjorken sum rule \cite{Bjorken:1966jh, Bjorken:1969mm}:
\begin{align}
\label{BJSum}
\int_0^1 \! dx  \left[g_1^p(x,k^2) - g_1^n(x,k^2)\right] =: \frac{g_A}{6}  \left[ 1 - \frac{\alpha_{g_1}(k^2) }{\pi} \right]\,,
\end{align}
where $g_1^{p,n}$ are the spin-dependent proton and neutron structure functions, whose extraction requires measurements using polarised targets, on a kinematic domain appropriate to deep inelastic scattering,
and $g_A$ is the nucleon isovector axial-charge.  Merits of this definition are outlined elsewhere \cite{Deur:2016tte}; and they include: the existence of data for a wide range of $k^2$ \cite{%
Deur:2005cf, Deur:2008rf, Deur:2014vea,
Ackerstaff:1997ws, Ackerstaff:1998ja, Airapetian:1998wi, Airapetian:2002rw, Airapetian:2006vy,
Kim:1998kia,
Alexakhin:2006oza, Alekseev:2010hc, Adolph:2015saz,
Anthony:1993uf, Abe:1994cp, Abe:1995mt, Abe:1995dc, Abe:1995rn, Anthony:1996mw, Abe:1997cx, Abe:1997qk, Abe:1997dp, Abe:1998wq, Anthony:1999py, Anthony:1999rm, Anthony:2000fn, Anthony:2002hy};
tight sum-rules constraints on the behaviour of the integral at the IR and UV extremes of $k^2$;
and the isospin non-singlet feature of the difference, which both ensures the absence of mixing between quark and gluon operators under evolution and suppresses contributions from numerous processes that are hard to compute and hence might obscure interpretation of the integral in terms of an effective charge.

The world's data on $\alpha_{g_1}(k^2)$ are depicted in Fig.\,\ref{Figwidehatalpha} and therein compared with our prediction for the RGI PI running-coupling $\hat{\alpha}(k^2)$.
As discussed in connection with Eq.\,\eqref{eq:PIpert}, all reasonable definitions of a QCD effective charge must agree on $k^2\gtrsim m_p^2$.  Our approach guarantees this connection, \emph{e.g}.\ in terms of the widely-used $\overline{\rm MS}$ running coupling \cite{Tanabashi:2018oca}:
\begin{subequations}
\label{AgreeCouplings}
 \begin{align}
 \hat{\alpha}(k^2) & = \alpha_{\overline{\rm MS}}(k^2) ( 1 + 1.09 \, \alpha_{\overline{\rm MS}}(k^2) + \ldots ) \,,\\
 \label{ag1MSbar}
 \alpha_{g_1}(k^2) & = \alpha_{\overline{\rm MS}}(k^2) ( 1 + 1.14 \, \alpha_{\overline{\rm MS}}(k^2) + \ldots ) \,,
 \end{align}
\end{subequations}
where Eq.\,\eqref{ag1MSbar} may be built from, \emph{e.g}.\ Refs.\,\cite{Kataev:1994gd, Baikov:2010je}.  Evidently, $\hat\alpha$ and $\alpha_{g_1}$ differ by
$\lesssim 0.5 \,\alpha_{\overline{\rm MS}}(k^2)$
on any domain within which perturbation theory is valid.

Significantly, there is also excellent agreement between $\hat\alpha$ and $\alpha_{g_1}$ on the IR domain, $k^2 \lesssim m_p^2$.  We attribute this to the isospin non-singlet character of the Bjorken sum rule, which ensures that contributions from many hard-to-compute processes are suppressed, and these same processes are absent from $\hat{\alpha}(k^2)$.

The RGI PI charge, $\hat\alpha(k^2)$, has been used in an exploratory calculation of the proton's elastic electromagnetic form factors in the hard-scattering regime \cite{Rodriguez-Quintero:2018wma}.  More recently, it has been employed to develop the kernel for DGLAP evolution \cite{Dokshitzer:1977sg, Gribov:1972, Lipatov:1974qm, Altarelli:1977zs} of the pion's parton distribution functions (PDFs) \cite{Ding:2019qlr, Ding:2019lwe}.  Here we provide a novel perspective on this latter application.

To establish a context, we recall that the first DSE applications to the calculation of hadron observables \cite{Maris:1997tm} (and many subsequent studies), chose a renormalisation scale deep in the spacelike region: $\zeta= 19\,$GeV, primarily to ensure simplicity in the nonperturbative renormalisation procedure.
This choice entails that the dressed quasiparticles obtained as DSE solutions remain intact and thus serve as the dominant degrees-of-freedom for all observables.  That is adequate for infrared quantities, such as hadron masses: flexibility of model parameters and the bridge with QCD enable valid predictions to be made.  However, it generates errors in form factors and parton distributions.  With form factors, the correct power-law behaviour is obtained, but the scaling violations deriving from anomalous operator dimensions are incorrect (see, \emph{e.g}.\ Ref.\,\cite{Maris:1998hc}); and for distributions, the natural connection between the renormalisation point and the reference scale for evolution equations is lost, again because parton loops are suppressed when renormalising a given truncated study at deep spacelike momenta so the computed anomalous dimensions are wrong.

The solution to these problems is to renormalise the DSE solutions at the hadronic scale, $\zeta_H\lesssim m_p$, where the dressed quasiparticles \emph{are} the correct degrees-of-freedom \cite{Raya:2016yuj, Gao:2017mmp, Chen:2018rwz, Ding:2018xwy, Ding:2019qlr, Ding:2019lwe, Xu:2019ilh}.  This recognises that a given meson's Poincar\'e covariant wave function and correlated vertices, too, must evolve with $\zeta$ \cite{Lepage:1979zb, Efremov:1979qk, Lepage:1980fj}.  Such evolution enables the dressed-quark and -antiquark degrees-of-freedom, in terms of which the wave function is expressed at a given scale $\zeta$, to split into less well-dressed partons via the addition of gluons and sea quarks in the manner prescribed by QCD dynamics.  These effects are automatically incorporated in bound-state problems when the complete quark-antiquark scattering kernel is used; but aspects are lost when that kernel is truncated.

Recognising this, the initial predictions for the pion PDFs in Refs.\,\cite{Ding:2019qlr, Ding:2019lwe} were presented at the scale $\zeta_H$, whereat the pion is purely a bound-state of a dressed-quark and dressed-antiquark; hence, sea and glue distributions are zero.  At any $\zeta>\zeta_H$, each distribution is then obtained via QCD evolution from these initial forms.  Here arises the question: ``What is the natural value of $\zeta_H$?'', something which has been asked in all studies since Ref.\,\cite{Jaffe:1980ti}.

Refs.\,\cite{Ding:2019qlr, Ding:2019lwe} provided an answer and prediction in terms of $\hat\alpha$.  Namely, capitalising upon the fact that QCD possesses a RGI PI effective charge, which saturates in the infrared owing to the dynamical generation of a gluon mass-scale, it introduced the simplified running coupling:
\begin{align}
\label{alphaPI}
\tilde\alpha(k^2) = \frac{4\pi}{\beta_0 \ln[(m_\alpha^2+k^2)/\Lambda_{\rm QCD}^2]}\,,
\end{align}
with $m_\alpha$ chosen to reproduce the known value of $\alpha_0$, Eq.\,\eqref{alpha0}.  Here, $m_\alpha\sim m_0$ serves as an essentially nonperturbative scale whose existence ensures that parton modes with $k^2 \lesssim m_\alpha^2$ are screened from interactions.  Thus, $m_\alpha$ marks the boundary between soft and hard physics; accordingly, Refs.\,\cite{Ding:2019qlr, Ding:2019lwe} identified
\begin{equation}
\label{setzetaH}
\zeta_H=m_\alpha
\end{equation}
and used Eq.\,\eqref{alphaPI} to define evolution of all PDF moments.

For example, with ${\mathpzc q}^\pi(x;\zeta_H)$ being the pion's valence-quark distribution function and
\begin{equation}
M_n(t) = \int_0^1 dx\,x^n\,{\mathpzc q}^\pi(x;\zeta(t))\,,
\end{equation}
$\zeta(t) = \zeta_H {\rm e}^{t/2}$, then Refs.\,\cite{Ding:2019qlr, Ding:2019lwe} expressed the PDF at any scale $\zeta>\zeta_H$ via the moments
\begin{equation}
\label{MellinEvolution}
M_n(t) = M_n(0) \, \exp\left[-\bar\gamma_0^n \int_0^t dz\,\frac{\tilde\alpha(\zeta^2(z))}{4\pi}\right]\,,
\end{equation}
$\bar\gamma_0^n= -(4/3) (3 + 2/(n + 1)/(n + 2) - 4 \sum_{i=1}^{n+1}(1/i))$.  This is leading-order QCD evolution based on $\tilde\alpha\approx \hat \alpha$.

Given that the perturbative expansion parameter in Eq.\,\eqref{MellinEvolution} is $\tilde\alpha(\zeta)/[4\pi] \lesssim 1/4$ $\forall \zeta>\zeta_H$, Refs.\,\cite{Ding:2019qlr, Ding:2019lwe} argued that leading-order evolution should provide a good approximation; and, in fact, working from this assumption, the resulting prediction for the pion's valence-quark PDF is in excellent agreement with existing data \cite{Conway:1989fs, Aicher:2010cb}.  Additionally, sound predictions for the pion's glue and sea-quark distributions were also obtained.

Such phenomenological successes support a broader view of $\hat\alpha$.  Namely, this RGI PI running coupling can also be interpreted as that special process-dependent effective charge for which evolution of the moments of all pion PDFs is defined by the one-loop formula expressed in terms of $\tilde\alpha\approx\hat\alpha$ \cite{Rodriguez-Quintero:2019fyc}.  It then unifies the Bjorken sum rule with pion (meson) PDFs.  From this perspective, the question of the size of an expansion parameter no longer arises.  Instead, introduction of $\hat \alpha$  into the leading-order formula for these observables serves to express the canonical all-orders resummed and infrared finite result for each quantity.

As remarked following Eq.\,\eqref{BJSum}, it has usually been thought that there is a different process-dependent charge for each observable.  However, it is now apparent that $\hat\alpha$, itself RGI and PI, unifies two very distinct sets of measurements, \emph{viz}.\ the pion's structure function and the leading moment of nucleon spin-dependent structure functions.  Hence, $\hat\alpha(k^2)$ is emerging as a good candidate for that object which truly represents the interaction strength in QCD at any given momentum scale \cite{Dokshitzer:1998nz}.

\section{Summary and Perspective}
\label{epilogue}
Using modern lattice-QCD configurations generated with three domain-wall fermions at the physical pion mass, we computed QCD's renormalisation-group-invariant (RGI) process-independent (PI) effective charge, $\hat\alpha(k^2)$ [Fig.\,\ref{Figwidehatalpha}], which, in being completely determined by the gluon two-point function, is a unique strong-interaction analogue of the Gell-Mann--Low effective coupling in QED.  Owing to the dynamical breakdown of scale invariance, expressed through emergence of a RGI gluon mass-scale, with calculated value $m_0/{\rm GeV} = 0.43(1)$ [Eq.\,\eqref{DeltaAnswers}], this running coupling saturates at infrared momenta: $\hat\alpha(k^2=0)/\pi=0.97(4)$  [Eq.\,\eqref{alpha0}].  Our results are parameter-free predictions.

The calculated RGI PI charge is smooth and monotonically decreasing on $k^2\geq 0$ and is known to unify a wide range of observables, \emph{inter alia}: hadron static properties \cite{Wang:2018kto, Qin:2019hgk, Xu:2019sns, Souza:2019ylx}; parton distribution amplitudes of light- and heavy-mesons \cite{Shi:2015esa, Ding:2015rkn, Chouika:2017rzs, Binosi:2018rht} and associated elastic and transition form factors \cite{Raya:2016yuj, Gao:2017mmp, Chen:2018rwz, Ding:2018xwy, Xu:2019ilh}.  In addition, $\hat\alpha(k^2)$ is:
(\emph{i}) pointwise (almost) identical to the process-dependent (PD) effective charge, $\alpha_{g_1}$, defined via the Bjorken sum rule;
(\emph{ii}) capable of marking the boundary between soft and hard physics; and
(\emph{iii}) that PD charge which, used at one-loop in the QCD evolution equations, delivers agreement between pion parton distribution functions calculated at the hadronic scale and experiment.
In playing so many diverse r\^oles, $\hat\alpha(k^2)$ emerges as a strong candidate for that object which properly represents the interaction strength in QCD at any given momentum scale.

Our study supports a conclusion that the Landau pole, a prominent feature of perturbation theory, is screened (eliminated) in QCD by the dynamical generation of a gluon mass-scale and the theory possesses an infrared stable fixed point.  Accordingly, with standard renormalisation theory ensuring that QCD's ultraviolet behaviour is under control, QCD emerges as a mathematically well-defined quantum field theory in four dimensions.

\begin{acknowledgments}
We are grateful for insights from L.~Chang, M.~Ding, J.~Pawlowski, J.-L.~Ping, K.~Raya, S.\,M.~Schmidt and H.-S.~Zong; and to the RBC/UKQCD collaboration, especially P.~Boyle, N.~Christ, Z.~Dong, C.~Jung, N.~Garron, B.~Mawhinney and O.~Witzel, for access to the lattice configurations employed herein.
Our calculations benefited from the following resources: CINES, GENCI, IDRIS (Project ID 52271); and the IN2P3 Computing Facility.
Work supported by:
National Natural Science Foundation of China (under Grant No.\,11805097),
Jiangsu Provincial Natural Science Foundation of China (under Grant No.\,BK20180323).
Jiangsu Province \emph{Hundred Talents Plan for Professionals};
Generalitat Valenciana, under grant Prometeo/2019/087;
Spanish Ministry of Economy and Competitiveness (MINECO), under grant no.\ FPA2017-84543-P.
\end{acknowledgments}


\providecommand{\newblock}{}

\end{document}